\documentclass[showpacs,preprintnumbers,amsmath,amssymb]{revtex4}
\usepackage{graphicx}
\usepackage{dcolumn}
\usepackage{bm}

\usepackage{color}

\begin{document}

\title{Plasmonic surface traps with arbitrary shape for cold atoms} 
\author{Matthias Mildner$^1$}
\author{Andreas Horrer$^2$}
\author{Monika Fleischer$^2$}
\author{Claus Zimmermann$^1$}
\author{Sebastian Slama$^1$}

\affiliation{Physikalisches Institut and Center for Collective Quantum Phenomena $^1$, resp. Institute for Applied Physics $^2$, both in LISA$^+$, Eberhard Karls Universit\"{a}t T\"ubingen, Auf der Morgenstelle 14, D-72076 T\"ubingen, Germany}

\date{\today}

\begin{abstract}
This paper reports on conceptual and experimental work towards the realization of plasmonic surface traps for cold atoms. The trapping mechanism is based on the combination of a repulsive and an attractive potential generated by evanescent light waves that are plasmonically enhanced. The strength of enhancement can be locally manipulated via the thickness of a metal nanolayer deposited on top of a dielectric substrate. Thus, in principle arbitrary potential landscapes can be generated. We present simulations of a plasmonic lattice potential using a gold grating with sinusoidally modulated thickness. Experimentally, a first plasmonic test structure is presented and characterized. Furthermore, the surface potential landscape is detected by reflecting ultracold atom clouds from the test structure revealing the influence of both evanescent waves. A parameter range is identified, where stable traps can be expected.

\end{abstract}


\maketitle

\section{Introduction}
The realization of surface traps for cold atoms has recently attracted considerable attention \cite{Vetsch10, Thompson13}. Cold atoms that are trapped at a nanoscale distance from solid surfaces can be strongly coupled to photonic bandgap nanoresonators and guided modes in optical nanofibers with applications as hybrid quantum devices, for instance as single atom switches \cite{Shea13}, and controlled phase gates \cite{Tiecke14}. Plasmonic surfaces have the advantage that light can be even stronger confined far below the diffraction limit with resulting ultra-high coupling constants. In this context the field of quantum plasmonics is emerging \cite{Tame13}. We have recently shown that the fluorescence of ultracold atoms can be coupled with high cooperativity to surface plasmon polaritons \cite{Stehle14}. In those experiments a large cloud of atoms was moving towards the surface. A surface trap provides much better control over the position of the atoms and avoids averaging effects over the distance from the surface. Surface traps in state-of-the-art experiments are based either on red-detuned dipole traps that arise from the reflection of a focused laser beam from the surface \cite{Thompson13}, or on combined laser fields with red and blue detuning, propagating along nanofibers as evanescent waves (EW) \cite{Vetsch10}, similar to early EW experiments on a glass prism \cite{Rychtarik04}. Here, we propose a mechanism that is based on the plasmonic enhancement of both red- and blue-detuned evanescent waves that are excited at the surface of a thin gold layer. By design of the local gold layer thickness the enhancement can be laterally manipulated across the surface. Thus, in principle arbitrary potential landscapes and trap shapes can be generated, for instance  lattice heterostructures including defects, two-dimensional patterns, and even Fibonacci lattices, extending the vision of recent proposals \cite{Gullans12}. Another scheme for the realization of surface traps with sub-wavelength optical lattices is published back-to-back \cite{bernon18}.\\ 

This paper explains the working principle of a surface trap based on plasmonically enhanced evanescent waves with different color. It analyzes in detail the influence of important experimental parameters, in particular of laser power and displacement between the evanescent waves. A parameter range is identified, in which stable traps can be formed. Moreover, the paper shows how the trap depth can be manipulated across the surface via the local thickness of a gold nanolayer: optical near-field simulations at a sinusoidal gold grating demonstrate the formation of a one-dimensional plasmonic lattice. A corresponding test structure with a period well below the optical wavelength has been fabricated and is characterized. Finally, the paper shows measurements of the surface potential generated at the test structure by the overlap of the evanescent waves. These measurements are done by reflecting ultracold atom clouds, i.e. by a method which has been developed within our group for the detection of surface potentials \cite{Bender10}. 

\section{Simulation of plasmonic two-color traps}

\subsection{Plasmonic trapping potential on a flat surface}
\begin{figure}[ht]
\centerline{\scalebox{0.8}{\includegraphics{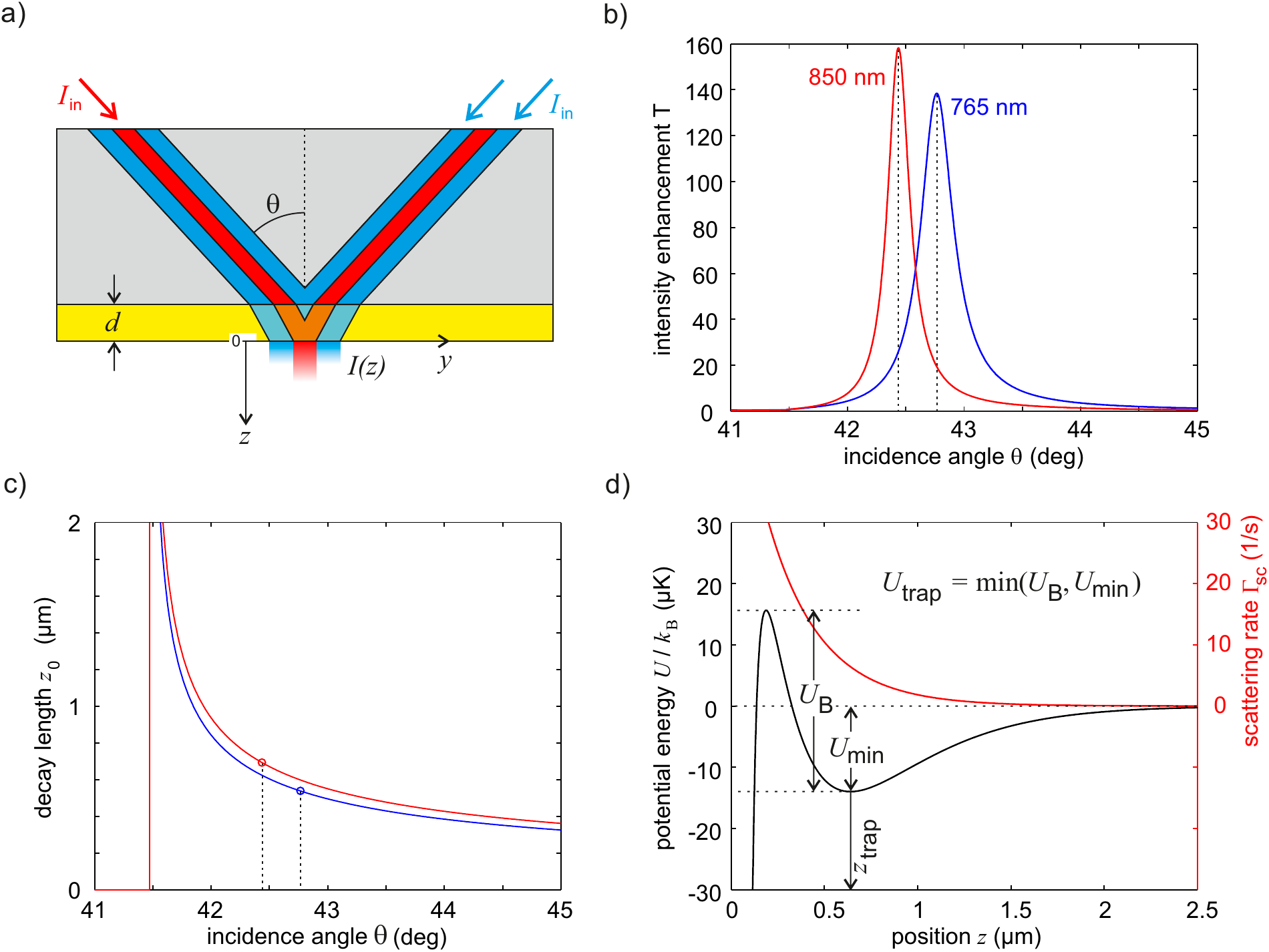}}}
\caption{a) Experimental setup: two laser beams with $\lambda_\mathrm{r}=850~\mathrm{nm}$ and $\lambda_\mathrm{b}=765~\mathrm{nm}$ are reflected by total internal reflection. They excite surface plasmon polaritons at the surface of a thin gold layer with $d=50~\mathrm{nm}$ thickness. b) Intensity enhancement $T=I_0/I_\mathrm{in}$ as function of incidence angle $\theta$ for both colors. c) Decay length $z_{0}$ as function of incidence angle $\theta$ for both colors. d) The attractive and repulsive potential corresponding to the two light fields sum up to a potential minimum $U_\mathrm{min}$. The trap is closed towards the surface via a potential barrier with height $U_\mathrm{B}$. The effective trap depth is thus limited by $U_\mathrm{trap}=\min\left(U_\mathrm{B}, U_\mathrm{min}\right)$. In this graph the incident laser intensities are chosen to be $I_\mathrm{in,\mathrm{red}}=439~\mathrm{W/cm^2}$ and $I_\mathrm{in,\mathrm{blue}}=289~\mathrm{W/cm^2}$. With these parameters a trap with trapping frequency $\omega_z=2\pi\times 26.4~\mathrm{kHz}$ and trap depth $U_\mathrm{trap}=k_B\times 14~\mathrm{\mu K}$ is generated, with Boltzmann constant $k_B$. The position of the minimum at $z_\mathrm{trap}=650~\mathrm{nm}$ is slightly shifted (not resolvable in the Figure) by the attractive Casimir-Polder potential that dominates at small distances, i.e. for $z\lesssim 100~\mathrm{nm}$. The trap position is also only little shifted by the linear gravitational potential which prevails at large distances. The rate at which photons are incoherently scattered from an arbitrary atom is $\Gamma_\mathrm{sc} < 10/s$ at the position of the trapping minimum.}
\label{fig:flat_surface}
\end{figure}

The principle of the proposed plasmonic trap is first demonstrated at an unstructured gold surface. The trap is generated by overlapping two evanescent waves with red / blue detuning giving rise to an attractive / repulsive surface potential, respectively. The potentials decay exponentially with distance $z$ from the surface as 
\begin{equation}
\label{eq:U_rb}
U_\mathrm{red/blue}=U_{0,\mathrm{red/blue}}\cdot \exp\left(-2\cdot\frac{z}{z_{0,\mathrm{red/blue}}}\right)~,
\end{equation}
with potential strength $U_{0,\mathrm{red/blue}}$ at the surface and decay length $z_{0,\mathrm{red/blue}}$ of the evanescent wave \cite{Stehle11}. The potential strength depends on the laser intensity $I_{0,\mathrm{red/blue}}$ at the surface and the detuning $\Delta_{1,2}=\omega-\omega_{D_\mathrm{1,2}}$, with laser frequency $\omega$ and atomic resonance frequencies $\omega_{D_\mathrm{1,2}}$ of the Rubidium D1- and D2-Line \cite{Grimm00}:
\begin{equation}\label{eq:U_0}
U_{0,\mathrm{red/blue}}= I_{0,\mathrm{red/blue}}\cdot\frac{\pi c_0^2 \Gamma}{2\omega^3}\left(\frac{2}{\Delta_2}+\frac{1}{\Delta_1}\right)~,
\end{equation}
with linewidth $\Gamma$ of the transitions and the speed of light $c_0$. Thus, for negative (red) detuning the potential attracts the atoms towards the surface, and for positive (blue) detuning, it repels them from the surface. In Kretschmann configuration \cite{Kretschmann68}, the laser beams are reflected by total internal reflection within the dielectric substrate on top of which a thin gold layer is fabricated, Fig.~\ref{fig:flat_surface}a). The intensities $I_{0,\mathrm{red/blue}}$ at the surface are proportional to the intensities $I_\mathrm{in,red/blue}$ of the incident laser beams inside the dielectric. They are analytically calculated with the transfer matrix method described in \cite{Deutsch95}. Due to the excitation of surface plasmon polaritons the intensity is resonantly enhanced at a certain incidence angle, i.e. at the plasmon angle $\theta=\theta_\mathrm{pl}$, Fig.~\ref{fig:flat_surface}b). Due to the dispersion of the gold layer, the plasmon angle depends on the wavelength $\lambda$. For gold, the plasmon angle of the red detuned light field with $\lambda_\mathrm{r}=850~\mathrm{nm}$ is smaller than that of the blue detuned light field with $\lambda_\mathrm{b}=765~\mathrm{nm}$. By tuning the two incidence angles to the values with largest enhancement, the decay length $z_0$ \cite{Grimm00} is larger for the case of the red-detuned EW, Fig.~\ref{fig:flat_surface}c). Thus, the attractive potential reaches out further from the surface than the repulsive one. For a suitable choice of laser powers the sum of both potentials generates a potential minimum. The exact position of this minimum is also influenced by the attractive Casimir-Polder (CP) potential 
\begin{equation}\label{eq:U_CP}
U_\mathrm{CP}= -\frac{C_4}{z^3(z+\lambda/2\pi)}~
\end{equation}
with transition wavelength $\lambda$ and $C_4$-coefficient $C_4=4.6\cdot 10^{-55}$  \cite{Stehle11} for Rubidium atoms at a gold surface. Furthermore, it is influenced by the gravitation 
\begin{equation}\label{eq:U_g}
U_\mathrm{g}= - mgz~,
\end{equation}
with atom mass $m$ and gravitational acceleration $g$. In the present setup the gravitational force points away from the surface. For typical experimental parameters the combined potential
\begin{equation}\label{eq:U_tot}
U_\mathrm{tot}= U_\mathrm{red} + U_\mathrm{blue} + U_\mathrm{CP}+ U_\mathrm{g}~,
\end{equation}
generates a trap as shown in Fig.~\ref{fig:flat_surface}d). The trap consists in the $z$-direction of a potential minimum at distance $z_\mathrm{trap}$ from the surface with potential depth $U_\mathrm{min}$. Between the potential minimum and the surface a potential barrier with height $U_\mathrm{B}$ is generated that protects the trap against atom loss towards the surface. For even shorter distances $z\lesssim 100~\mathrm{nm}$ beyond the barrier the strongly attractive Casimir-Polder potential dominates and accelerates the atoms towards the surface.

\subsection{Influence of laser power}
\begin{figure}[ht]
\centerline{\scalebox{0.8}{\includegraphics{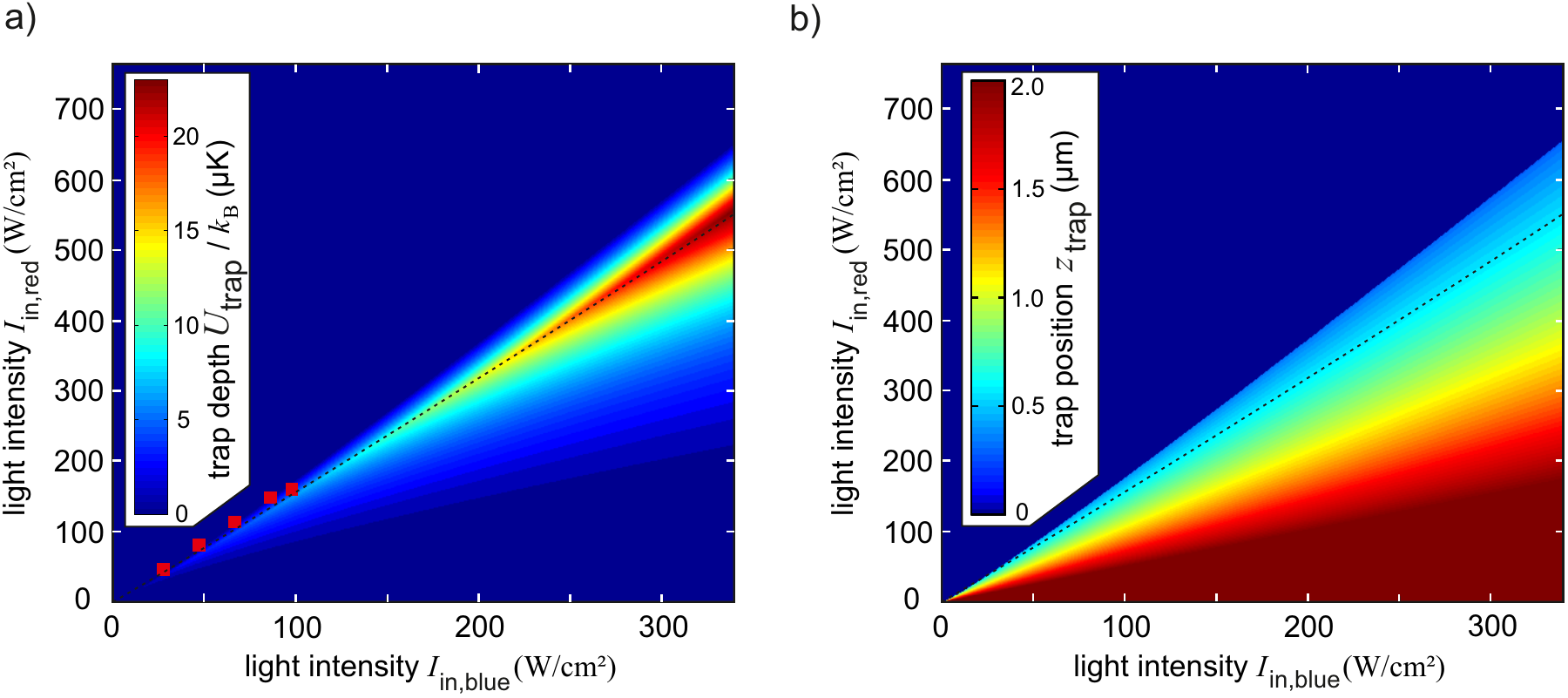}}}
\caption{a) Trap depth $U_\mathrm{trap}$ and b) trap distance $z_\mathrm{trap}$ as functions of the laser intensities $I_\mathrm{in,red/blue}$. The plotted range of intensities corresponds to the maximum laser power available in the present experiment of $P_\mathrm{red}=160~\mathrm{mW}$ and $P_\mathrm{blue}=160~\mathrm{mW}$. The dashed line indicates the intensity ratios with maximum trap depth. Additionally, experimentally measured data points (red squares) which correspond to the situation where half of the atoms are reflected from the surface for increasing intensity $I_\mathrm{in,red}$, determined from Fig.~\ref{fig:measurement_potential}c), are plotted in a).}
\label{fig:laser_power}
\end{figure}

The shape of the potential depends critically on the intensities of both the red- and the blue-detuned evanescent wave. First of all, the formation of a trap requires the repulsive potential to be strong enough in order to partially compensate the attractive Casimir-Polder potential and to generate a potential barrier that repels the atoms from the surface. The stronger the intensity is, the closer this barrier is formed to the surface \cite{Bender10}. A practical upper limit for the intensity is given by heating of the atom cloud due to incoherent scattering of photons from the laser. Ultimately, the maximum power that can be used is limited by the fact that the gold layer is destroyed due to laser-induced heating. A trap can be formed at distances larger than that of the barrier, if the attractive potential dominates the repulsive one in this distance range.  Thus, a minimum intensity of the red-detuned laser field is required to generate a trap. Above this minimum the trap depth increases with the laser intensity, see Fig.~\ref{fig:laser_power}a). However, if the intensity $I_\mathrm{in,red}$ is too large for a fixed $I_\mathrm{in,blue}$, the attractive potential dominates the repulsive one also at the distance of the barrier, which reduces the barrier height and ultimately opens the trapping potential towards the surface. This reduces the effective trap depth. In Fig.~\ref{fig:laser_power}a) the trap depth is plotted against the intensities of both lasers, revealing the parameter range where stable traps can be expected. The trap depth is maximum along the dashed diagonal line. Below this line the barrier height is larger than the minimum $U_\mathrm{B}>U_\mathrm{min}$. On the diagonal line barrier height and potential minimum are equal $U_\mathrm{B}=U_\mathrm{min}$, and above the line the barrier height is smaller the minimum $U_\mathrm{B}<U_\mathrm{min}$.
The corresponding trap distance from the surface is plotted in Fig.~\ref{fig:laser_power}b). It is noteworthy that along the dashed line, the distance of the trap from the surface is almost constant at a value of $z_\mathrm{trap}\approx 550~\mathrm{nm}$.

\subsection{Influence of beam profile and displacement}
\begin{figure}[ht]
\centerline{\scalebox{0.8}{\includegraphics{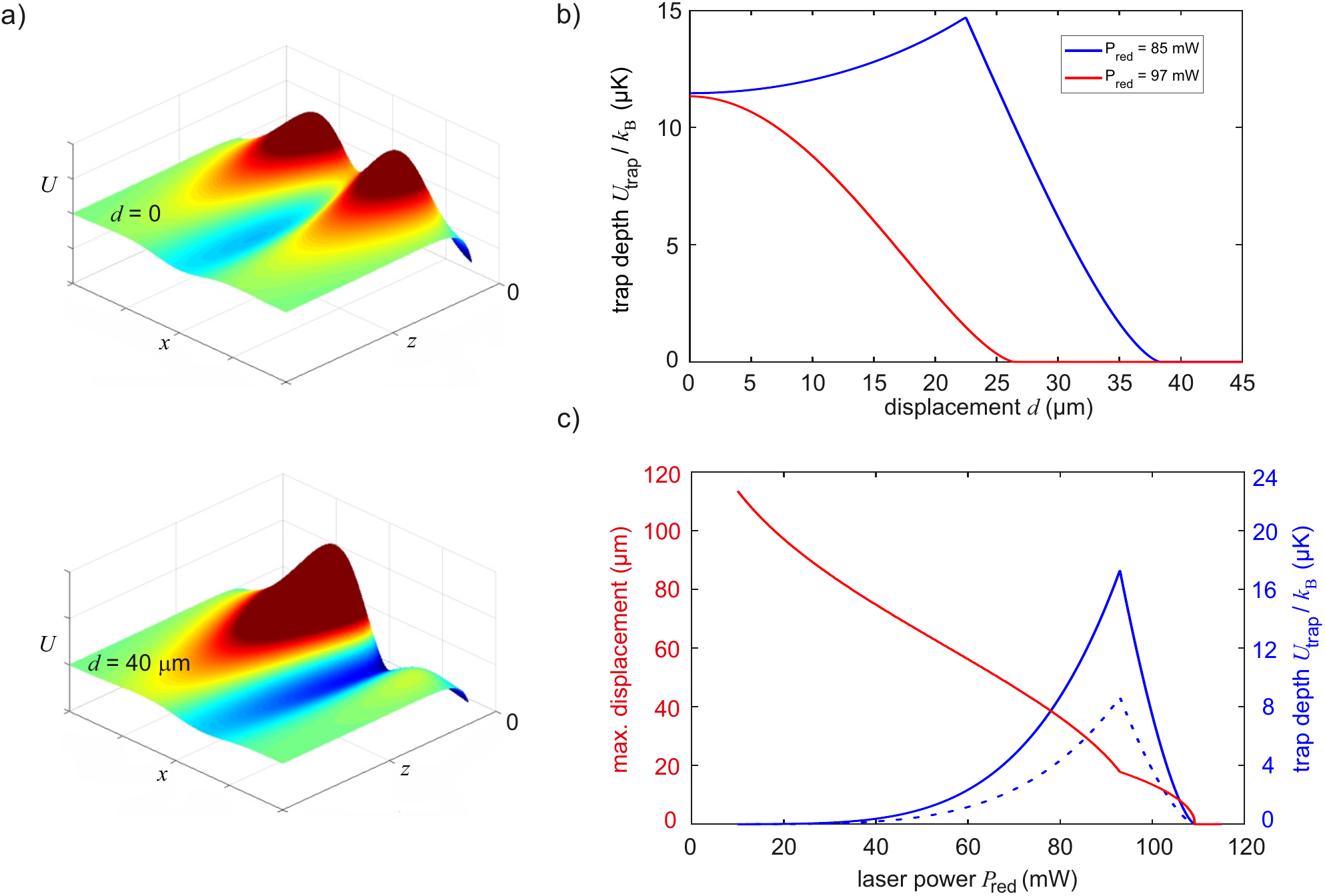}}}
\caption{a) A lateral displacement $d$ between the center of the red- and the blue-detuned evanescent waves can lead to holes in the trapping potential towards the surface. b) The functional dependence of the trap depth on the displacement depends on the laser power $P_\mathrm{red}$, here exemplarily plotted for $P_\mathrm{blue}=130~\mathrm{mW}$. The maximum displacement is defined at the point where the trap depth reaches zero. c) The maximum displacement decreases with increasing laser power $P_\mathrm{red}$, whereas the trap depth in the displaced trap (solid line) is increasing until a maximum is reached. The dashed line corresponds to the trap depth without displacement, for comparison. The beam waists in this simulation are $w_\mathrm{0,red,x}=100~\mu\mathrm{m}$, $w_\mathrm{0,red,y}=135~\mu\mathrm{m}$, $w_\mathrm{0,blue,x}=150~\mu\mathrm{m}$, $w_\mathrm{0,blue,y}=204~\mu\mathrm{m}$. The beam waists are larger in the $y$-direction for both beams due to the projection onto the surface at incidence angles of $\theta_\mathrm{red}=42,44^\circ$, and $\theta_\mathrm{blue}=42,76^\circ$.}
\label{fig:lateral_offset}
\end{figure}	
The laser fields used in the experiment have a typically Gaussian transverse beam profile. The intensity at the surface is thus given by
\begin{equation}
I(x,y)= I_0\cdot \exp\left(-\frac{2x^2}{w_{0,x}^2}-\frac{2y^2}{w_{0,y}^2}\right)~,
\end{equation} 
with beam waists $w_{0,x}$ and $w_{0,y}$. This profile has a strong influence on the potential shape in the radial direction, as both intensities are reduced with radial distance from the centers of the beams. A trap that is confined also in the radial direction requires a ratio $I_\mathrm{in,red}/I_\mathrm{in,blue}$ that is in the right parameter range for all radial positions, i.e. the ratio should - in the best case - stay close to the diagonal dashed line in Fig.~\ref{fig:laser_power}a). This requires that the beam waist of the red-detuned EW $w_\mathrm{0,red}$ should not be larger than that of the blue-detuned EW $w_\mathrm{0,blue}$. Otherwise, the trap is opened towards the surface at some radial distance from the beam centers. For practical reasons $w_\mathrm{0,red}$ is chosen to be even smaller than $w_\mathrm{0,blue}$. The trap may also be opened when the centers of the two beams are not aligned with each other, see Fig.~\ref{fig:lateral_offset}~a). The dependence of the trap depth on the displacement in shown in Fig.~\ref{fig:lateral_offset}~b). For sufficiently small laser power and small displacement (blue line) the trap depth is initially increasing due to the fact that the maximum intensity ratio $I_\mathrm{in,red}/I_\mathrm{in,blue}$ approaches the dashed line in Fig.~\ref{fig:laser_power}a). For larger displacement the trap depth decreases again and ultimately reaches zero, thus defining a maximum displacement. For large laser power corresponding to an intensity ratio above the dashed line in Fig.~\ref{fig:laser_power}a) the trap depth decreases immediately to zero (red line). The maximum displacement is shown in Fig.~\ref{fig:lateral_offset}~c) together with the trap depth as a function of the laser power $P_\mathrm{red}$. For increasing laser power $P_\mathrm{red}$ the maximum displacement is decreasing whereas the trap depth is increasing until its maximum is reached. Thus, a compromise has to be made between a large trap depth and a large range of tolerable misalignment.

\subsection{Plasmonic lattice potential}
\begin{figure}[ht]
\centerline{\scalebox{0.8}{\includegraphics{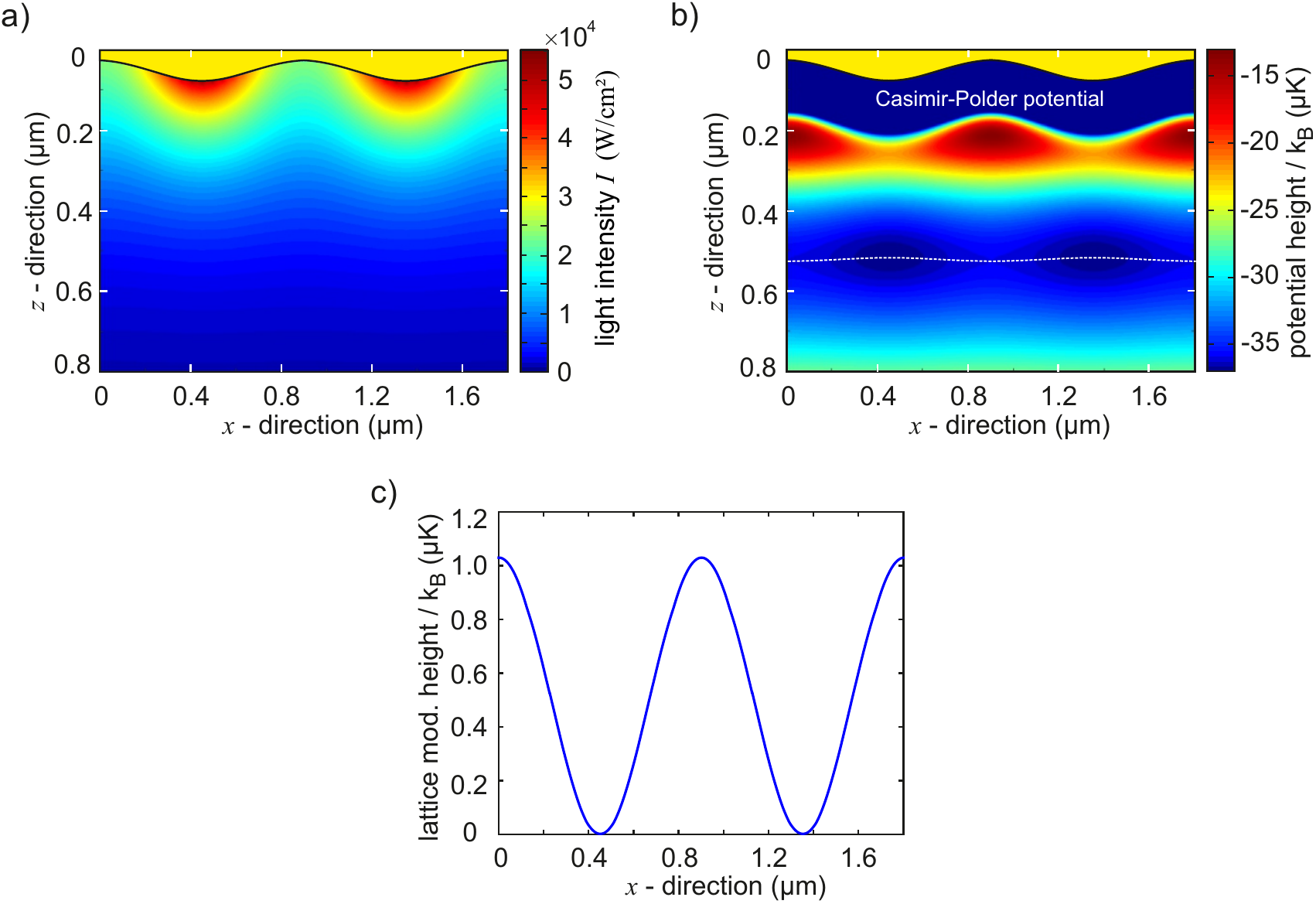}}}
\caption{a) Light intensity distribution (Comsol) at the sinusoidal gold grating for a lattice constant of $a=900~\mathrm{nm}$.
b) Potential landscape above the sinusoidal gold grating. For this simulation a lattice constant of $a=900~\mathrm{nm}$, average layer thickness $d_m=50~\mathrm{nm}$, and modulation depth $d_m=25~\mathrm{nm}$ are chosen. The incoming light intensities are $I_\mathrm{in,red}=731~\mathrm{W/cm^2}$ and  $I_\mathrm{in,blue}=430~\mathrm{W/cm^2}$. Thus, a trap is formed at a distance of $z_\mathrm{trap}\approx 500~\mathrm{nm}$ with potential minimum $U_\mathrm{min}=k_B\times 37~\mathrm{\mu K}$ and a trap depth $U_\mathrm{trap}=k_B\times 15~\mathrm{\mu K}$ limited by the barrier to the surface. The trapping frequency is $\omega_z=2\pi\times 31~\mathrm{kHz}$. At distances from the surface shorter than $\sim 200~\mathrm{nm}$ the strongly attractive Casimir-Polder potential shifts the potential height to values outside the plotted range (dark blue region). In the $x$-direction, the potential height is modulated and forms an optical lattice. c) Modulation of the lattice potential with reference set to $U_\mathrm{min}$. The barrier height between neighboring lattice sites is on the order of $E_B\approx k_B\times 1~\mathrm{\mu K}$, the trapping frequency in the $x$-direction is $\omega_x=2\pi\times 730~\mathrm{Hz}$.}
\label{fig:potential_landscape}
\end{figure}	
The potentials drawn in Fig.~\ref{fig:flat_surface} are valid for a constant gold layer thickness. In the following a lateral variation in the layer thickness $d(x)$ in the $x$ direction is taken into account. In order to describe the resulting intensity variation $I(x)$ of the evanescent waves correctly, it is not sufficient to calculate the intensity locally, i.e. as if the layer thickness was constant with a value corresponding to $d(x)$. In particular, the shape of the surface plays an important role for the enhancement. In \cite{Stehle11} we showed that a step grating where the layer thickness varies abruptly between zero and a value of $d=50~\mathrm{nm}$ looses the feature of enhancing the EW intensity when the lattice constant is reduced. In order to avoid this reduction of enhancement, we consider now a smooth, sinusoidal modulation of the layer thickness along the $x$-direction. The gold layer thickness is thus given by
\begin{equation}\label{eq:d(x)}
d(x)= d_m + d_0\cdot \sin\left(\frac{2\pi}{a}\cdot x\right)~,
\end{equation}
with periodicity $a$ of the grating and modulation depth $d_0$ around an average layer thickness $d_m$. The intensity landscape cannot be calculated analytically like on a flat surface, instead it is here calculated using COMSOL Multiphysics with the radio frequency-module (RF-module). The model is solved by the finite element method. Is is composed as a block with height $3000~\mathrm{nm}$ ($z$-direction), depth $10~\mathrm{nm}$ ($y$-direction), and width $900~\mathrm{nm}$ ($x$-direction), corresponding to the chosen lattice constant of the sinusoidal structure. The block is divided in $z$ - direction in four different layers. The first layer with a thickness of $1000~\mathrm{nm}$ corresponds to the glass substrate with a refractive index of $n=1.51$. The second layer simulates the sinusoidal gold structure on the glass substrate, and the third and fourth layer with $1000~\mathrm{nm}$ thickness each correspond to vacuum, respectively to an additional Perfectly Matched Layer (PML) in order to damp the transmitted wave and avoid artificial reflection. The electromagnetic wave is simulated as being irradiated from the glass side at the bottom of the model under an inclination $\theta$ with respect to the $y$-direction, see also Fig.\ref{fig:flat_surface}. Scattering boundary conditions are chosen in order to make this lower boundary transparent to reflected plane waves and to avoid artificial multiple reflections within the glass layer. In order to simulate the whole problem using a limited model size, periodic boundary conditions are chosen for the $x$- and $y$-direction. In $x$-direction the periodicity is given by the $q=2\pi/a$-vector of the lattice and in $y$-direction it is given by the $k_y=2\pi/\left(\lambda\cdot\cos(\theta)\right)$-vector of the incident field. The Comsol simulations reveal that the local enhancement depends on the lattice constant. For lattice constants of $a\lesssim 5~\mathrm{\mu m}$ the intensity at the hills is larger than for a plane surface with equal thickness. The simulated intensity landscape $I(x,z)$ is plotted in Fig.~\ref{fig:potential_landscape}~a) for a grating period of $a=900~\mathrm{nm}$, average layer thickness $d_m=50~\mathrm{nm}$, modulation depth $d_m=25~\mathrm{nm}$, and an optical wavelength of $\lambda=765~\mathrm{nm}$. Above the hills of the grating the enhancement is increased compared to a plane layer with equal thickness. Above the valleys of the grating it is reduced. This variation of enhancement across the grating generates the lattice potential. 
The total potential landscape shown in Fig.~\ref{fig:potential_landscape}~b) is calculated from the simulated intensities via (\ref{eq:U_tot}). A potential minimum is formed at a distance of $z_\mathrm{trap}\approx 500~\mathrm{nm}$. Due to the sinusoidal shape of the surface, the position $z_\mathrm{trap}(x)$ of the trap varies slightly ( $\lesssim 10~\mathrm{nm}$), depending on the $x$-position. The potential height along this sinuous line, i.e. approximately along the $x$-direction, is modulated such that an optical lattice is formed, see Fig.~\ref{fig:potential_landscape}~c). The barrier height between the lattice sites can be adjusted by the intensity of the red laser. For typical experimentally achievable parameters it is on the order of $2\%-3\%$ of the trap depth. An important feature to note is that the intensity modulation of the EWs along $x$ is substantial only at a distance from the surface smaller than the grating period. For larger distances the modulation smears out. Thus, the optical lattice can be generated only at distances from the surface smaller than the lattice period.  

\section{Fabrication and characterization of plasmonic sinusoidal grating}
\begin{figure}[ht]
\centerline{\scalebox{0.8}{\includegraphics{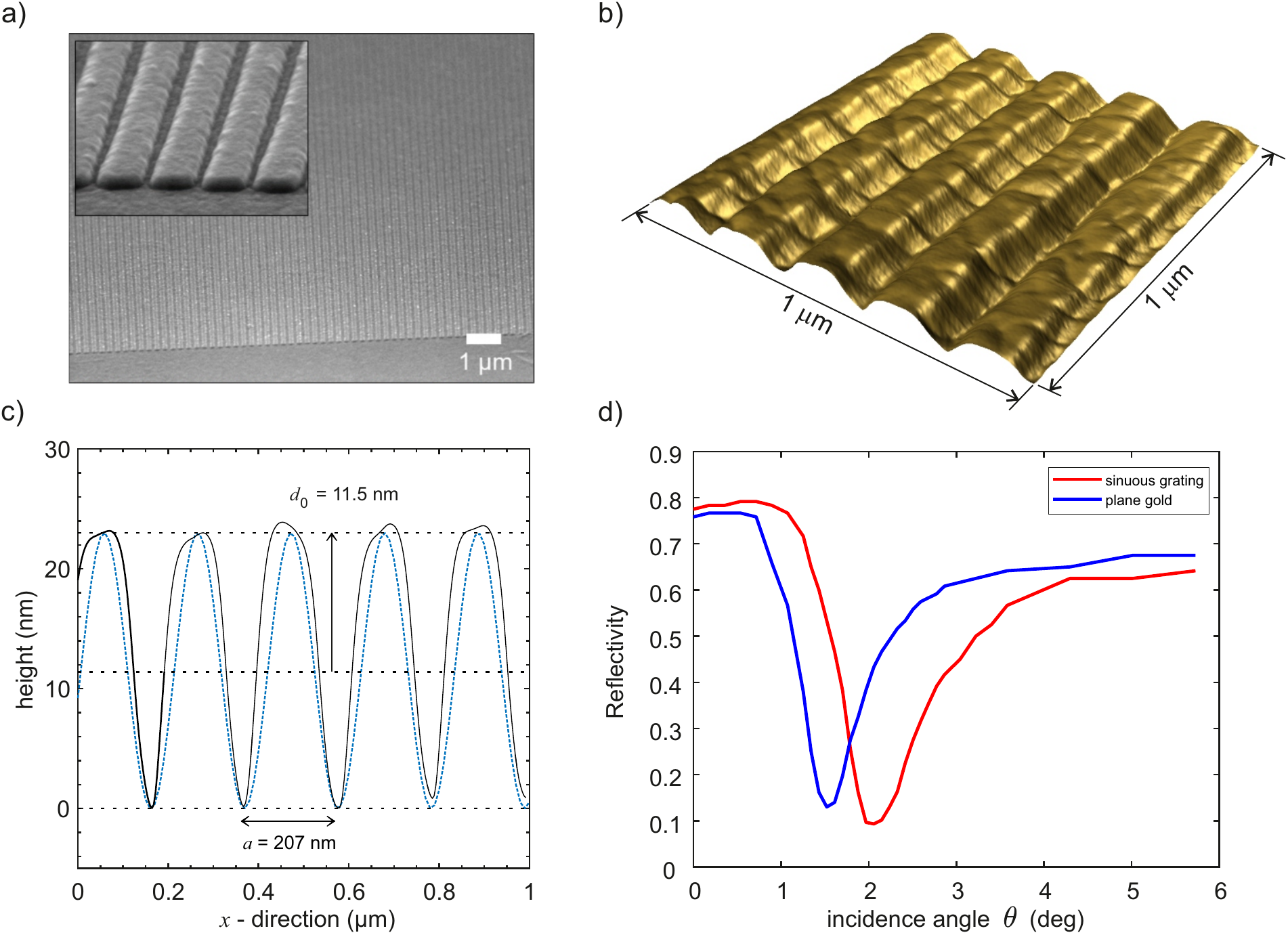}}}
\caption{a) SEM image after annealing: overview and close-up view (inset: $1~\mathrm{\mu m}$), b) AFM surface scan, and c) averaged line scan (black line) with fit of a sine curve (dashed blue line). d) Reflection measurement on a plane gold surface(blue) and on the sinusoidal grating (red). The incidence angle $\theta$ is measured in this figure with respect to the angle of total reflection.}
\label{fig:fabrication}
\end{figure}	
In order to demonstrate the applicability of the proposed method a sample with a plasmonic grating consisting of a gold layer with approximately sinusoidally varying thickness has been fabricated. The lattice constant has been designed as $a=200~\mathrm{nm}$, i.e. considerably smaller than in the simulations shown. Thus, on the one hand these structures will not lead to a substantial modulation of the lattice potential, on the other hand, the sample demonstrates that the limits in the miniaturization imposed by the nanofabrication process are well below the simulated structure size. The fabrication process includes three major steps: (i) optical lithography in order to form gold pads on a glass sample, (ii) electron beam lithography (EBL) in order to form ridges on top of the pads, and (iii) an annealing step in order to smoothen and reshape the surface. At first, 3 x 3 square pads with sizes of $(200~\mathrm{\mu m})^2$ (for the gratings) as well as a large $(1~\mathrm{mm}^2)$ pad (for a plane reference surface) were patterned in a layer of resist on a glass substrate by optical lithography. A gold film with a nominal height of $30~\mathrm{nm}$ was deposited by thermal evaporation. The resist was lifted off in acetone, and the gold pads remained on the glass surface. In the second step, the substrate was spin-coated with a $180~\mathrm{nm}$ thick layer of polymethyl methacrylate (PMMA) resist, which was subsequently covered with a layer of aluminum with a height of $10~\mathrm{nm}$ by thermal evaporation. The purpose of the aluminum was to render the sample conductive for EBL. One-dimensional line arrays with a nominal period of $a=200~\mathrm{nm}$ were aligned to the underlying pads and exposed. The aluminum was removed by etching in an alkaline solution (ma-D 331 developer, MicroChem), and the PMMA was developed in isopropanol and methyl isobutyl ketone with a mixing ratio of 3:1 for $70~\mathrm{s}$. Then, a gold film with a nominal height of $20~\mathrm{nm}$ was deposited as before. After lift-off of the PMMA, ridges with a roughly rectangular profile and a width of about $150~\mathrm{nm}$ had formed on the gold pads. The sample was annealed for $60~\mathrm{min}$ at $200~\mathrm{^\circ C}$ on a hotplate, which resulted in a rounding of the edges of the ridges to an approximately sinusoidal shape, and a decrease of the surface roughness.
The surface was investigated using scanning electron microscopy (SEM) and atomic force microscopy (AFM). An exemplary SEM image can be seen in Figure \ref{fig:fabrication}a). It shows the edge of the area covered with gold ridges on a gold pad viewed under a tilting angle of $70^\circ$, illustrating the homogeneity of the lattice over a large area on the surface. The inset shows an area with a width of one micron at higher magnification. An AFM scan over an area of  $1~\mathrm{\mu m}$ x $1~\mathrm{\mu m}$ of the gold surface can be seen in Figure \ref{fig:fabrication}b). A roughness of about $\mathrm{1-2}~\mathrm{nm}$ could be estimated from the scan. A surface profile, which was obtained by averaging over several line scans perpendicular to the ridges of the surface, is shown in Figure \ref{fig:fabrication}c). A sine curve was fitted to the dips in the profile. In its broader upper part, the profile deviates from the sine shape. The period of the sine curve was measured to $a=207~\mathrm{nm}$ and its amplitude to $d_{0}=11.5~\mathrm{nm}$. The $1~\mathrm{mm}$-thick glass substrate was glued index-matched on a plane dove prism, and surface plasmon polaritons were excited by illuminating the surface in Kretschmann-configuration with p-polarised light \cite{Kretschmann68}. Fig.~\ref{fig:fabrication}d) shows a measurement of the light (wavelength $780~\mathrm{nm}$) reflected from the plasmonic surfaces as a function of the incidence angle $\theta$. Clear minima in reflection are observed at the so-called plasmon angle, corresponding to the excitation of surface plasmons. The blue curve was measured at the plane gold surface for reference, whereas the red curve was measured using one of the nine sinusoidal gratings. Apart from a shift of the plasmon angle, which can be attributed to humidity in the grooves of the grating, surface plasmons can be excited equally well in both structures. The shift disappeared after installing the structures into the experimental vacuum chamber and baking. The slight difference in the contrast and the width of the resonances can be explained by a variation in the gold layer thickness between the two samples. A repeated reflection measurement after several months of experiments with cold atoms (not shown) revealed that the reflectivity at the plasmon angle increased from $9\%$ up to $27\%$ at the grating. This deterioration of the plasmon excitation may be caused either by overheating of the surface or by reacting of rubidium with the gold grating.\\

\section{Detection of plasmonic surface potentials}
\begin{figure}[ht]
\centerline{\scalebox{0.8}{\includegraphics{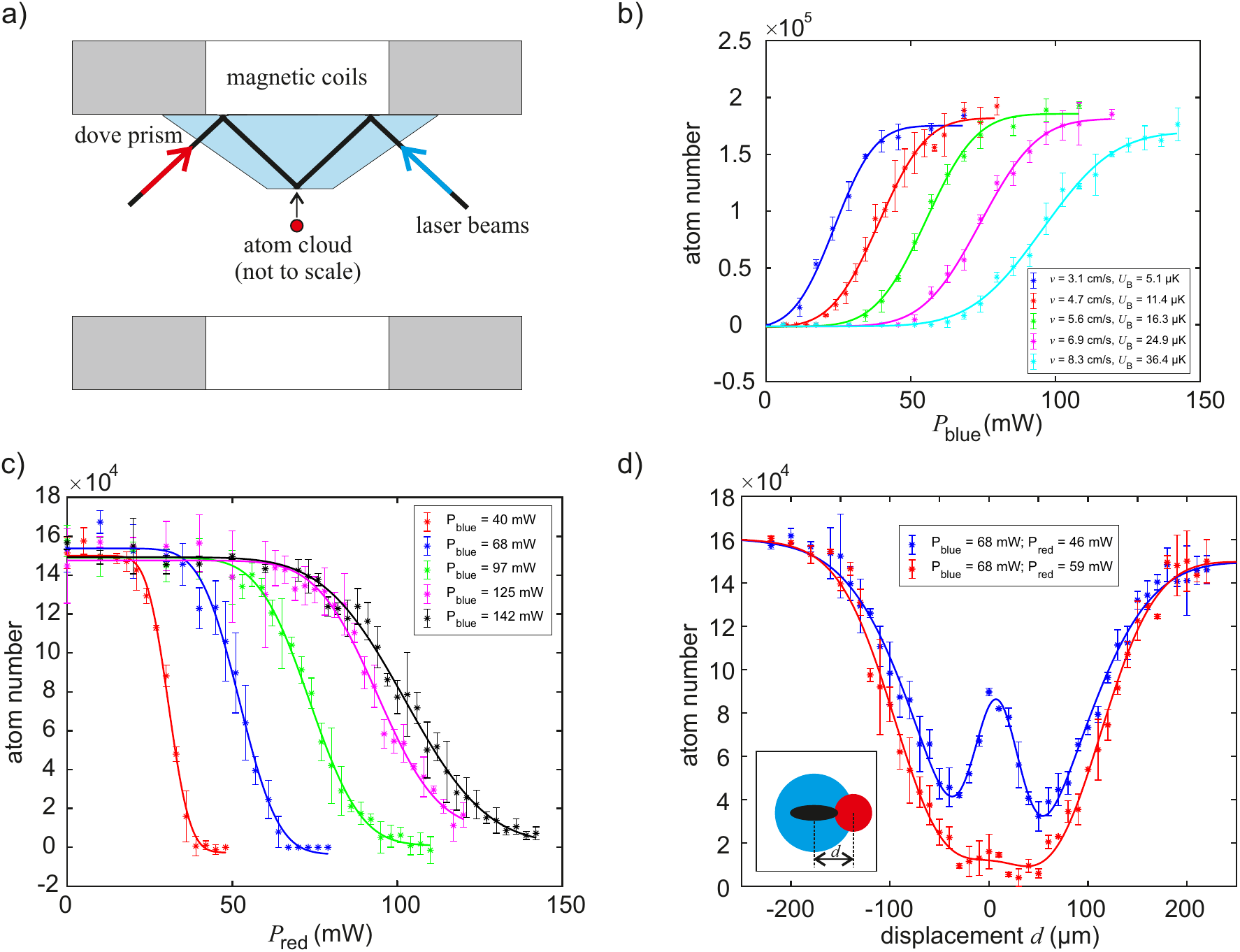}}}
\caption{a) Sketch of experimental setup including the magnetic coils and the prism. b) Reflection of atom clouds from the potential barrier generated by the blue-detuned evanescent wave alone. c) Due to the contribution of the attractive potential the number of reflected atoms decreases when the power of the red-detuned laser beam is increased. d) The red-detuned laser beam is laterally shifted from the center of the blue-detuned laser beam and from the position of the atom cloud (inset). This leads to an additional loss of atoms through holes in the potential. The solid lines are a guide to the eye only. The larger the power of the red-detuned laser beam is, the smaller is the maximum displacement. All plotted error bars are due to statistical errors.}
\label{fig:measurement_potential}
\end{figure}

In order to experimentally test the idea of generating plasmonic surface traps and lattices using evanescent waves with different colors, we detect the corresponding potential landscape by reflection of ultracold atom clouds from the surface. For that purpose the dove prism including the fabricated plasmonic structures has been installed into an ultracold-atom-experiment. We present here only the major steps of the experiment, details can be found in \cite{Bender09, Bender10, Stehle14}. A cigar-shaped cloud of rubidium atoms (with $1/e^2$-radii $r_x\sim 120~\mu\mathrm{m}, r_y\sim 22~\mu\mathrm{m}, r_z\sim 22~\mu\mathrm{m}$) is prepared in a magnetic trap between two magnetic coils, see Fig.~\ref{fig:measurement_potential}a). The cloud is cooled by forced radio-frequency evaporation to a temperature of $T=1~\mu\mathrm{K}$. The dove prism is placed in between the coils such that the atom cloud can be moved towards the surface and also laterally across it using external magnetic fields. The two blue- and red-detuned laser beams are irradiated into the prism from counter propagating directions. Both lasers are power stabilized by a PID-feedback loop with $0.1\%$ stability and are guided in optical fibers to the vacuum chamber. The position of the beams on the prism surface can be controlled via a translation stage with an accuracy of $10~\mathrm{\mu m}$. The laser beams are focused to beam waists of $\omega_b=146\pm5~\mathrm{\mu m}$ and $\omega_r=90\pm5~\mathrm{\mu m}$ at the position of the prism surface. The incidence angle of both beams onto the surface is controlled by laser mirrors outside the vacuum chamber and is tuned to the plasmon angle for each beam.\\

In a first experiment the barrier height that is generated by the blue-detuned evanescent wave alone is measured by reflecting cold atom clouds with variable velocity from the surface. The barrier height is measured via the loss of reflected atoms which occurs as soon as the kinetic energy of the atoms exceeds the height of the barrier. In this case the atoms pass the barrier and are accelerated towards the surface by the Casimir-Polder potential where they heat up to room temperature and partially adsorb. The experiment is carried out in the following way \cite{Bender09, Bender10}. Atom clouds are accelerated towards the surface by a sudden displacement of the magnetic trapping potential. The velocity is tuned by the chosen displacement. After acceleration of the cloud the magnetic trap is switched-off and a magnetic gradient field is switched-on in order to compensate the gravitational potential. Thus, the atoms move towards the surface with almost constant velocity where they are partially reflected depending on the irradiated laser power and the atomic velocity. The number of reflected atoms is measured by absorption imaging after variable time of flight (typically $15~\mathrm{ms}$). Typical reflection measurements are shown in Fig.~\ref{fig:measurement_potential}b). For each value of atomic velocity the number of reflected atoms is measured for increasing laser power. The number rises from zero to a maximum in a power range where the barrier height is comparable with the kinetic energy of the atoms. The width of this rise is determined by the atomic velocity spread. The value of the barrier height is calculated from the point where half of the atoms are reflected. This procedure connects the barrier height with the laser power resp. its intensity. Thus, the enhancement of the evanescent wave is determined to $I_\mathrm{0,b}/I_\mathrm{in,b}=45$. This value is $33\%$ of the maximum theoretical value. This discrepancy from the theoretical value can be attributed to several loss processes like angle deviation from the plasmon angle, polarization deviation from pure p-polarization, layer thickness uncertainty, and surface roughness. This first experiment proves that the evanescent wave is plasmonically enhanced. \\

A second experiment investigates the influence of the red-detuned evanescent wave on the barrier height. The atom cloud is now adiabatically moved within the magnetic trap towards the surface and -- after interaction with the surface -- back away from it. This is repeated for various values of $P_\mathrm{blue}$ and $P_\mathrm{red}$. The minimum distance of the atoms from the surface in this process is adjusted such that all atoms are lost if both lasers are switched off. This guarantees that all atoms interact with the surface. If only the blue-detuned laser beam is switched on, a considerable part of the atoms is reflected. Increasing the power $P_\mathrm{red}$ for constant $P_\mathrm{blue}$ adds an attractive contribution to the surface potential and eventually reduces the barrier height. Thus, the number of reflected atoms is decreasing, as can be seen in Fig.~\ref{fig:measurement_potential}c). The intensity pairs corresponding to the value of $P_\mathrm{blue}$ and $P_\mathrm{red}$ at which half of the maximum number of atoms are reflected are inserted into Fig.~\ref{fig:laser_power}a) as red squares. The plotted intensities take into account that the plasmonic enhancement is reduced to $33\%$ for both fields, as measured from Fig.~\ref{fig:measurement_potential}b). The plotted data points lie slightly above the line where the trap depth is maximum. This is reasonable, as the barrier height towards the surface is reduced above this line. The fact that the data points follow the slope of the line indicates that the actual potential shape coincides with the theoretical prediction.\\

A third experiment investigates the precision with which the two laser beams have to be overlapped in order to generate stable traps. Again, the atom cloud is moved to the surface and back away from it within the magnetic trap. The powers $P_\mathrm{red}$ and $P_\mathrm{blue}$ are now kept fixed, while the red-detuned laser beam is laterally displaced by a variable distance $d$ from the center of the blue-detuned laser beam. The position where the atom cloud approaches the surface has been adjusted to the center of the blue-detuned laser beam and is kept fixed in this experiment. As shown in Fig.~\ref{fig:lateral_offset}a) a lateral displacement of the two beams leads to holes in the trapping potential. Thus, the number of reflected atoms decreases, as can be seen in the measurements shown in Fig.~\ref{fig:measurement_potential}d) for  $P_\mathrm{red}=46~\mathrm{mW}$ and $P_\mathrm{blue}=68~\mathrm{mW}$ (blue data). The number of reflected atoms reaches a minimum for a displacement of $d\approx 50~\mu\mathrm{m}$. For even larger displacements the reflected atom number rises again. In this situation the displacement is larger than the extension of the atom cloud in the magnetic trap and the atoms are no longer influenced by the attractive potential. A second set of measured data with $P_\mathrm{red}=59~\mathrm{mW}$ and $P_\mathrm{blue}=68~\mathrm{mW}$ (red data) does not show a substantial increase of atom numbers for $d=0$. In this situation the barrier height is reduced already for perfectly overlapped beams such that no stable trap is generated.

\section{Conclusion}
This paper presents a method for the generation of plasmonic surface traps and lattices based on the combination of blue- and red-detuned evanescent waves that are plasmonically enhanced above a thin gold nanolayer. Simulations of the potential landscape reveal that trap depths $U_\mathrm{trap}> k_B\times 10~\mathrm{\mu K}$ with trap frequencies $\omega_z\sim 2\pi\times \textrm{several }10~\mathrm{kHz}$, $\omega_r\sim 2\pi\times 150~\mathrm{Hz}$, and incoherent light scattering rates $\Gamma_\mathrm{sc} < 10/s$ can be obtained. These values are compatible with ultracold atom experiments. The traps are formed at a typical distance from the surface of $z_\mathrm{trap}\sim 500~\mathrm{nm}$. Finite-element simulations show that plasmonic lattices can be generated by a sinusoidal manipulation of the gold nanolayer. With modulation along one direction (i.e. the $x$-direction) the lattice consists of a one-dimensional array of elongated traps with lattice barrier heights on the order of $k_B\times 1~\mathrm{\mu K}$ and trapping frequencies on the order of $\omega_z\sim 2\pi\times \textrm{several }10~\mathrm{kHz}$, $\omega_x\sim 2\pi\times 1~\mathrm{kHz}$, and $\omega_y\sim 2\pi\times 150~\mathrm{Hz}$, at a lattice constant of $a=900~\mathrm{nm}$. The presented method is applicable in principle to arbitrary potential shapes including lattice heterostructures, two-dimensional patterns and even Fibonacci lattices. Lattices with periods below the optical wavelength would require dramatically shorter trap distances from the surface which brings the here presented method to its limits. Shorter trap distances might however be reached by adding standing optical waves that are reflected from the surface. We will analyze this possibility in future work.\\

Experimentally, a first test structure has been fabricated and characterized, albeit the grating period of $a=200~\mathrm{nm}$ is too small to generate a substantial lattice modulation at the given trap distance. Reflecting laser beams from the gold surface in Kretschmann configuration showed that surface plasmons can be excited on the gold grating with comparable efficiency as on a flat gold surface. Ultracold atom clouds have been reflected from the gold grating demonstrating the effect of both the blue- and the red-detuned evanescent wave and the impact of a displacement between the two beams. The observed behavior of the atomic reflectivity coincides with the theoretical prediction and approves the validity of the simulated potential landscape including the traps. The next step we will address in the future is the loading of the plasmonic traps with cold atoms. This is not a trivial task because the traps are rather steep compared to magnetic traps. Thus, it is not possible to fill the traps by a simple adiabatic transfer from the magnetic trap. It is rather necessary to provide an additional cooling process that is able to increase the atomic density within the surface trap. Moreover, the trap depth is not very deep such that only few atoms ($N\lesssim 1000$) can be trapped simultaneously. Thus, the question arises how to detect such low atom numbers, in particular in the face of stray light from the surface which may prevent standard in-situ fluorescence or absorption imaging. We are presently working on a dispersive detection scheme using near-resonant light that is reflected from the gold surface in Kretschmann configuration, similar to surface plasmon resonance sensors \cite{Homola99}. 

\section{Acknowledgement}
This work was performed in the context of the European COST Action MP1302 Nanospectroscopy and the European COST Action MP1403 Nanoscale Quantum Optics.

\section{Competing financial interests}
Competing financial interests do not exist.

\end{document}